\documentclass[12pt]{article}
\title{Reconstruction of N=1 supersymmetry from topological symmetry}
\usepackage{mathrsfs}
\usepackage{amsmath}

\global\arraycolsep=1pt
\usepackage[colorlinks=true,backref=true,linkcolor=black,anchorcolor=black,citecolor=black,filecolor=black,menucolor=black,pagecolor=black,urlcolor=black]{hyperref}
\usepackage[Symbol]{upgreek}
\usepackage{bbm}
\usepackage{dsfont}
\usepackage{amssymb}
\usepackage{textcomp}

\newcommand{\Scal}[1]{\biggl ({#1} \biggr )}
\newcommand{\scal}[1]{\bigl ({#1} \bigr )}
\def\bea{\begin{eqnarray}}
\def\eea{\end{eqnarray}}
\def\be{\begin{equation}}
\def\ee{\end{equation}}
\newcommand{\CR}{\nonumber \\*}

\newcommand{\trace}{\hbox {Tr}~}

\newcommand{\gra}[2]{{\scriptscriptstyle (#1 , #2 )}}

\def\L{{\cal L}}

\def\Lc{\mathscr{L}}

\def\bA{\overset{\circ}{A}}

\def\deltac{{\delta_{\bar c}}}

\def\t{\tilde}

\def\aalpha{{\dot{\alpha}}}
\def\bbeta{{\dot{\beta}}}

\usepackage[colorlinks=true,backref=true,linkcolor=black,anchorcolor=black,citecolor=black,filecolor=black,menucolor=black,pagecolor=black,urlcolor=black]{hyperref}

\usepackage[Symbol]{upgreek}
\usepackage{caption}
\usepackage{bbm}
\usepackage{dsfont}
\usepackage{amssymb}
\usepackage{textcomp}
\def\bea{\begin{eqnarray}}
\def\eea{\end{eqnarray}}
\def\be{\begin{equation}}
\def\ee{\end{equation}}

\def\L{{\cal L}}

\def\Lc{\mathscr{L}}

\def\bA{{\overset{\circ}{A}}}

\def\t{\tilde}

\def\aalpha{{\dot{\alpha}}}
\def\bbeta{{\dot{\beta}}}
\def\deltac{{\delta_{\bar c}}}

\begin{document}
\allowdisplaybreaks[1]
\renewcommand{\thefootnote}{\fnsymbol{footnote}}

\begin{titlepage}
\begin{flushright}
 LPTHE-05\\
\end{flushright}
\begin{center}
{{\Large \bf Reconstruction of $\mathcal{N}=1$ Supersymmetry\\ From Topological Symmetry }}
\lineskip .75em
\vskip 3em
\normalsize
{\large L. Baulieu\footnote{email address: baulieu@lpthe.jussieu.fr},
 G. Bossard\footnote{email address: bossard@lpthe.jussieu.fr},
 }\\ 
$^{*\dagger}$ {\it LPTHE, CNRS and Universit\'es Paris VI - Paris VII, Paris,
France}\footnote{
4 place Jussieu, F-75252 Paris Cedex 05, France.} 
\\
$^{* }$\it Department of Physics, Rockefeller University\footnote{
1230 York Avenue, New York, NY 10021,~USA.,~USA.} 
\\

\vskip 1 em
\end{center}
\vskip 1 em
\begin{abstract}
The scalar and vector topological Yang--Mills symmetries on
Calabi--Yau manifolds geometrically define consistent sectors of
Yang--Mills $D=4,6\ \mathcal{N}=1$ supersymmetry,
which fully determine the supersymmetric actions up to twist. For a
$CY_2$ manifold, both 
$\mathcal{N}=1,D=4$ Wess and Zumino and 
superYang--Mills theory can be reconstructed in this way. A superpotential can be introduced for the
matter sector, as well as the Fayet--Iliopoulos mechanism. For a
$CY_3$ manifold, the $\mathcal{N}=1, D=6$
Yang--Mills theory is also obtained, in a twisted form. Putting these results together
with those already known for the $ D=4,8$ $\mathcal{N}=2$ cases, we conclude that all Yang--Mills supersymmetries with 4,
8 and 16 generators are determined from topological symmetry on
special manifolds. 
\end{abstract}

\end{titlepage}
\renewcommand{\thefootnote}{\arabic{footnote}}
\setcounter{footnote}{0}



\renewcommand{\thefootnote}{\arabic{footnote}}
\setcounter{footnote}{0}



\section{Introduction}
In a recent paper \cite{BBT}, it was shown that the scalar and 
vectorial topological Yang--Mills symmetries can be directly
constructed, in four and eight dimensions, leading one to a
geometrical definition of a closed off-shell twisted sector of
Yang--Mills supersymmetric theories,  with 8 and 16 generators,
respectively. In fact, 
both scalar and vectorial topological symmetries completely determine
the supersymmetric theory, (up to a twist that exists on special
manifolds). Basically, the vector symmetry arises when
one associates reparametrization symmetry and topological symmetry. It
is important to work on manifolds that contain at least one
covariantly constant vector. The $\mathcal{N}=2$ Poincar\'e
supersymmetry is reached by untwisting the theory in the limit of
flat manifolds. 
 
 The possibility of directly twisting the $\mathcal{N}=1$
 super Yang--Mills theory in a ``microscopic'' TQFT was studied in
 \cite{jaco,Witten,thesis,dijk,BS}. 
In fact, the full topological symmetry $ s A_\mu=\Psi_\mu+D_\mu c$
 involves topological ghosts and antighosts, with twice as many degrees
 of freedom as there are in the gauge field, so it leads one to
 $\mathcal{N}=2$ supersymmetry. 
 To get the $\mathcal{N}=1$ supersymmetry algebra in a twisted form, the number
 of independent topological transformations must be reduced by half.
 This leads one to build a TQFT on a K\"{a}hler manifold, such that the
 gauge field can be splitted in holomorphic and antiholomorphic
 components, $A_1=A_{\gra{0}{1}}+A_{\gra{1}{0}}$. Only one of the
 components of $A$ undergoes topological transformations, with
 \cite{jaco,Witten,thesis,dijk,BS}: 
 \be \label{simple}
 s A_m=\Psi_m+D_m c\quad \quad s A_{\bar m}= D_{\bar m} c
 \ee
This holomorphic symmetry can be interpreted on a Calabi--Yau manifold as the symmetry of
classical actions, which couple forms $ B_{\gra{0}{n-2}}$, with only 
antiholomorphic components, to a Yang--Mills curvature $F=dA+A_{\,
  \wedge} A$ \cite{lastBT}: 
 \be 
 I_{2n}=
 \int_{M_{2n}} \Omega_{\gra{n}{0}\,\, \wedge} \trace B_{\gra{0}{n-2}\,\,
 \wedge} F_{\gra{0}{2}} 
 \ee
 The BRST-invariant gauge-fixing of such
 actions provides in a twisted way the $\mathcal{N}=1$ Wess and Zumino and 
 Yang--Mills models in 4 and 6 dimensions, on Calabi--Yau manifolds ~\cite{lastBT}.

Here we show that both scalar and vector topological BRST
symmetries can be also geometrically built for the $\mathcal{N}=1$
supersymmetry, in a way that justifies the choice of topological
gauge functions of \cite{lastBT}. In fact, the various properties of $\mathcal{N}=1$ supersymmetry can be reformulated, in the context of topological symmetry. 
 
 Let us briefly summarize the situation for getting the scalar and vector topological symmetry, and, eventually 
 $\mathcal{N}=2$ supersymmetry in twisted way. \cite{BBT}
 shows that, for special manifolds with dimensions 4 or 8 that contain at least
 one constant vector $\kappa$, one can define an
 extended horizontality condition with its Bianchi identity, which involves 
 the (twisted) fields of $\mathcal{N}=2$ supersymmetry in 4 and 8 dimensions. It reads:
\begin{gather} 
( d + s + \delta - i_{\kappa} ) \scal{ A + c + |\kappa| \bar c } + \scal{
 A + c + |\kappa| \bar c }^2
= F + \Psi + g(\kappa) \eta + i_{\kappa }
\chi + \Phi + |\kappa|^2 \bar\Phi 
\CR
( d + s + \delta - i_{\kappa } ) \scal{F + \Psi + g(\kappa)
 \eta + i_{\kappa } \chi + \Phi + |\kappa|^2 \bar\Phi }
\hspace{30mm}\CR \hspace{25mm}
 + \,[
A + c +|\kappa| \bar c \,,\, F + \Psi + g(\kappa) \eta + i_{\kappa } \chi +
\Phi + |\kappa|^2 \bar\Phi ] = 0
\label{nero}
\end{gather}
By expansion in form degree and ghost number, both equations
define the action of $s$ and $\delta$ on all the fields, with 
 the closure relations\footnote{$ \bA$ is a background connection that must be introduced for the
sake of global consistency, but can be chosen equal to zero for
trivial vacua.}:
 \begin{gather}
 s^2 = \delta^2 =0\hspace{10mm} \{s, \delta\} = \L_\kappa +
\delta_{\mathrm{gauge}}(i_\kappa \bA) 
\end{gather}
 The property that $s$ and $\delta$ close off-shell on
 a reparametrization, $ \{s, \delta\} = \L_\kappa $, is at the
heart of the property that the commutator of two supersymmetries is
a translation in the super Poincar\'e algebra. 
In flat space, one defines the twisted scalar supersymmetric operator
$Q = s_c$ and the vector supersymmetric operator
$Q_\mu$ from the equivariant vector operator $\deltac=
\kappa^\mu Q_\mu$, so one has $Q_\mu Q_\nu+Q_\nu Q_\mu= 2 g_{\mu\nu}
\delta_{\rm gauge} (\bar \Phi)$
and $ Q_\mu Q+ Q Q_\mu= D_\mu$.
We refer to \cite{BBT} for a detailed explanation of these formula and
the twisted fields that they involve, with their relationship with
$D=4, 8 \ \mathcal{N}=2$ Yang--Mills supersymmetry, and the way reparametrization symmetry is encoded in $s$ and $\delta$.

The aim of this paper is to understand the way the extended
 horizontality condition (\ref{nero}) applies to the case of Calabi--Yau
 manifolds, for reconstructing $\mathcal{N}=1$ supersymmetry. In fact, by separation of holomorphic and antiholomorphic sectors, $\mathcal{N}=1$ supersymmetry will appear. The relevant information on the Wess and Zumino and
Yang--Mills independent multiplets of $D=4\ {\rm or}\ D=6 \
\mathcal{N}=1 $ will be obtained as off-shell closed sectors 
 of the supersymmetry transformation laws. More precisely, 
the supersymmetry transformations will be encoded 
into scalar and vector
topological BRST symmetries, corresponding to 3 (resp. 4) twisted generators in 4
(resp. 6) dimensions. The topological construction has the great advantage of
purely geometrically determining
 a closed sector of the supersymmetric
algebra, which is large enough to completely determine the theory. Furthermore, it determines the Faddeev--Popov ghosts for the supersymmetry algebra, in a way that is relevant for a control of the covariant gauge-fixing of the Yang--Mills symmetry.


\section{Holomorphic vector symmetry in four dimensions}
\subsection{Pure Yang--Mills}
\label{yang}
We begin from the ``semi-horizontality'' condition for a Yang--Mills field $A$ and its Faddeev--Popov ghost $c$, on a 
K\"{a}hler 2-fold:
\be
\label{simpled}
( d + s ) \scal{ A + c} + \scal{ A +c}^2 = F + \Psi, 
\ee
 $\Psi= \Psi_m dz^m$ is the holomorphic 
 $1$-form topological ghost. If $J$ is the complex structure on the manifold, one has 
$
J \Psi = i \Psi
$, Eq.~(\ref{simpled}) reproduces the ``heterotic'' BRST
transformations, Eq.~(\ref{simple}), and includes the ghost dependence, with: 
\be \begin{array}{rclcrcl}
s A_m &=& \Psi_m + D_m c &\hspace{10mm}& s A_{\bar m} &=& D_{\bar m} c
\CR
s \Psi_m &=& -[c, \Psi_m] & & s c &=& -c^2 
\end{array} \ee
 The Euclidean vector ghost $\Psi_m{\scriptstyle
  (z^{\scriptscriptstyle m}, z^{\scriptscriptstyle \bar m})}$ must be
 considered as a complex field that counts for 2 real 
degrees of freedom in the quantum theory, as it will be explained
in section \ref{untwist}. 
To introduce the vector symmetry, we 
suppose that the manifold contains at least a covariantly
constant antiholomorphic vector $\kappa^{\bar m}$. $\kappa $ 
 defines the 
holomorphic $1$-form $g(\kappa)=g_{m\bar m}\kappa^{\bar m }d z ^{ m }
$. The norm of $\kappa$ is 
$ |\kappa|^2 = \kappa^{\bar m }{\bar \kappa _{\bar m}}=
i_{\bar \kappa} g(\kappa) $, where 
$\bar \kappa$ is the complex conjugate of $ \kappa$. 
The differential $\delta$ will be geometrically constructed, and is somehow the mirror of $s$. In flat space, formula must be expanded as series in $\kappa^{\bar
 m}$. The vector symmetry operator $Q_{\bar m}$ is then defined by the identification 
$\delta + |\kappa|\delta_{\rm gauge}( \bar c) =\kappa^{\bar m} Q_{\bar
 m}$, and  determines, together with $s + \delta_{\rm
 gauge}(c)$, the relevant closed sector of $\mathcal{N}=1$ supersymmetry. 

The dual of the ``holomorphic'' $1$-form topological ghost 
$\Psi_m$ is made of a pair of a scalar $\upeta$ and an
``antiholomorphic'' $2$-form $\, \upchi_{\bar m\bar n}$, counting
altogether for 2 real degrees of freedom. $\bar c$  is the
Faddeev--Popov antighost. 
The ghost anti-ghost dependent horizontality condition that defines
both $s$ and $\delta$ symmetry is: 
\be\label{hor}
( d + s + \delta - i_\kappa ) \scal{ A + c + {|\kappa |\bar c}} +
\scal{ A + c +{|\kappa |\bar c}}^2 = F + \Psi + g(\kappa) \upeta +
i_\kappa \, \upchi
\ee
This gives
\begin{gather} 
\begin{array}{rclcrcl}
s A + d_A c &=& \Psi &\hspace{10mm}& \delta A + d_A {|\kappa |\bar c}
&=& g(\kappa) \upeta + i_\kappa \, \upchi \CR 
s c + c^2 &=& 0 & & \delta{|\kappa |\bar c} + \scal{{|\kappa |\bar c}}^2
&=& 0 \end{array} \CR
\delta c + s |\kappa |\bar c + [c , {|\kappa |\bar c}] = i_\kappa
\scal{ A - \bA} 
\end{gather}
One defines the ``equivariant''
 differential 
 $s_c\equiv s + \delta_{
\mathrm{gauge}}(c)$ and $ \deltac \equiv \delta +
|\kappa | \delta_{\mathrm{gauge}}(\bar c)$.

The property $( d + s + \delta - i_\kappa )^2=0$,
is equivalent to 
 the Bianchi identity 
\be\label{horr} 
(d_A + s_c + \deltac - i_\kappa ) \scal{ F + \Psi + g(\kappa) \upeta +
 i_\kappa \, \upchi} = 0 
\ee
The introduction of two scalar fields, $b$ and $h$ allows one to remove the 
indeterminacy that occurs in the determination of $s\bar c$ and $s
\upeta $. Eq.~(\ref{hor}) and its
Bianchi identity (\ref{horr}) provide by expansion in ghost number and form degree the following BRST transformations: 
\bea &\begin{array}{rclcrcl}\label{final}
s_c A_m &=& \Psi_m &\hspace{10mm}& \deltac A_m &=& \kappa_m \upeta \\*
s_c A_{\bar m} &=& 0 & & \deltac A_{\bar m} &=& \kappa^{\bar n}
\, \upchi_{\bar n \bar m} \\*
s_c \Psi_m &=& 0 & & \deltac \Psi_m &=& \kappa^{\bar n} F_{\bar n m} -
\kappa_m h \\*
s_c \upeta &=& h & & \deltac \upeta &=& 0 \\*
s_c h &=& 0 & & \deltac h &=& \kappa^{\bar m } D_{\bar m } \upeta \\*
s_c \, \upchi_{\bar m \bar n} &=& F_{\bar m \bar n} & & \deltac \,
\upchi_{\bar m 
 \bar n} &=& 0
\end{array}& \\
\CR
&\begin{array}{rclcrcl}
s c &=& - c^2 &\hspace{1cm}& \delta c &=& \kappa^{\bar m} \scal{ A_{\bar
 m}- \bA_{\bar m}} - {|\kappa |b} \\*
s { \bar c} &=& b - [c, \bar c]& & \delta \bar c
 &=& - |\kappa| \,{\bar c}^2 \\*
s b &=& - [ c, b] & & \delta b &=& \kappa^{\bar n}
D_{\bar n} {\bar c} 
\end{array}&
\eea
By construction, one has the required relations:
\begin{gather}
s^2 = 0 \hspace{10mm} \delta^2 = 0\hspace{10mm} \{s , \delta\} = \L_\kappa +
\delta_{\mathrm{gauge}}(i_\kappa \bA) 
\end{gather}
One has also ``equivariant'' commutation relations for all fields, but 
$c$, $\bar c$ and $b$:
 \begin{gather}
 s_c^2 
 = 0\hspace{10mm}
 \deltac^2 = 0
\hspace{10mm}
 \{s_c , \deltac\} = \L_\kappa
+ \delta_{\mathrm{gauge}}(i_\kappa A)
 \end{gather}

The most general $\delta$-closed topological gauge function, which has ghost number $-1$, is gauge invariant and gives $\kappa$ independent renormalizable terms, is: 
\be
\Uppsi_{\mathrm{YM}} = \int_M d^4 x \sqrt{g} \, \trace\Scal{\, \frac{1}{2} \upchi^{mn} F_{mn} + \upeta \scal{h + iJ^{m\bar n} F_{m\bar n}}}
\ee
It defines the ungauge-fixed $s_c$ and $\delta_{\bar c}$ invariant action $I_{\mathrm{YM}}=s \Uppsi_{\mathrm{YM}} $. Integrating out the field $h$ gives:
\be
I_{\mathrm{YM}}= \approx \int_M d^4 x \sqrt{g} \, \trace\Scal{\frac{1}{2} F^{mn} F_{mn} + \frac{1}{4} \scal{ J^{m\bar n} F_{m \bar n}}^2 - \, \upchi^{mn} D_m
 \Psi_n + \upeta D^m \Psi_m }
\ee
This is the twisted form of the $\mathcal{N}=1$ supersymmetric
Yang--Mills action, as expressed in \cite{jaco}.

This invariant action is in fact $s\delta$-exact. One has indeed:
\be
\Uppsi_{\mathrm{YM}} = { \textstyle \frac{1}{(\kappa \cdot \bar \kappa)}} \, \delta \int_M d^4 x \sqrt{g} \,
\trace \biggl( \,\,\,\upeta \bar
 \kappa^m \Psi_m + \bar \kappa^{[m} g^{n]\bar m} \scal{
 3\, A_{[\bar m} \partial_m A_{n]} + 2 A_{[\bar m} A_m A_{n]}} \biggr)
\ee
The last term is nothing but the Chern--Simon
term 
$g(\bar \kappa) \scal{ A d A + \frac{2}{3} A^3}
$
\subsection{Wess and Zumino matter multiplet}
The matter multiplet  is defined from horizontality
conditions, for both f extended
fields $\bar \psi _{\bar m} dz^{\bar m} + \phi$ and $\frac{1}{2} \bar \chi_{mn}dz^m_{\, \wedge} dz^n + \kappa_m dz^m \bar \phi$. 
$\bar \chi$ is a ``holomorphic'' $2$-form, 
$\bar \psi$ a ``antiholomorphic'' $1$-form, and
$\phi$ and 
$\bar \phi$   two scalars. These fields are valued in an arbitrarily
given gauge group representation. The horizontality conditions and
Bianchi identities are: 
 \bea
(\bar\partial_A + s_c + \deltac - i_\kappa) \scal{\bar \psi + \phi}
&=& \bar\partial_A \bar \psi + i_\kappa B \CR
(d_A + s_c + \deltac - i_\kappa) \scal{\bar \chi+ g(\kappa) \bar \phi} &=&
\bar\partial_A \bar \chi+ T - g(\kappa) \scal{ \bar\partial_A \bar\phi + \bar \eta}\CR
(\bar\partial_A + s_c + \deltac - i_\kappa) \scal{\bar\partial_A \bar \psi +
 i_\kappa B} 
&=& \scal{F_{\scriptscriptstyle (0,2)} + i_\kappa \, \upchi}\scal{\bar \psi
 + \phi} \\*
(d_A + s_c + \deltac - i_\kappa) \scal{\bar\partial_A \bar \chi+ T -
 g(\kappa) \scal{ \bar\partial_A \bar\phi + \bar \eta}} &=& \scal{F + \Psi +
 g(\kappa) \upeta + i_\kappa \, \upchi} \scal{ \bar \chi+ g(\kappa)
 \bar\phi} \nonumber
\eea
 The gauge field $A$ obeys the same equations as defined in the
 previous section. This gives the following BRST transformations: 
\be \begin{array}{rclcrcl}
s_c \bar \psi_{\bar m} &=& -D_{\bar m} \phi &\hspace{10mm}& \deltac
\bar \psi_{\bar m} &=& \kappa^{\bar n} B_{\bar n \bar m} \\*
s_c \phi &=& 0 & & \deltac \phi &=& - \kappa^{\bar m} \bar \psi_{\bar m} \\*
s_c B_{\bar m \bar n} &=& 2 D_{[\bar m} \bar \psi_{\bar n]} + \,
\upchi_{\bar m \bar n} \phi & & \deltac B_{\bar m \bar n} &=& 0 \\*
s_c \bar \chi_{mn} &=& T_{mn} & & \deltac \bar \chi_{mn} &=& 2 \kappa_{[m} D_{n]}
\bar \phi \\*
s_c T_{mn} &=& 0 & & \deltac T_{mn} &=& \kappa^{\bar p} D_{\bar p}
\bar \chi_{mn} - 2 \kappa_{[m} D_{n]} \bar \eta- 2 \kappa_{[m} \Psi_{n]} \bar
\phi \\*
s_c \bar \phi &=& \bar \eta& & \deltac \bar \phi &=& 0\\*
s_c \bar \eta&=& 0 & & \deltac \bar \eta&=& \kappa^{\bar m} D_{\bar m} \bar\phi
\end{array} \ee
They represent the twisted scalar and vector supersymmetric transformations for a Wess and Zumino multiplet.

For a general simple non Abelian gauge group, the $\delta$ invariance uniquely determines the most general
renormalizable
and $\kappa$ independent 
 topological gauge function with ghost number -1 $,
 \Uppsi_{\mathrm{Matter}}$, as follows\footnote{We do not write
 explicitly the sum over the index of the 
 matter representation, so that, one has for instance $
 \phi \bar \phi \equiv \sum_A \phi_A \bar \phi_A $}:
\be
I_{\mathrm{Matter}}= s\Uppsi_{\mathrm{Matter}}, \quad \quad 
\Uppsi_{\mathrm{Matter}} = \int_M d^4 x \sqrt{g} \, \Scal{
 \frac{1}{2} \bar \chi^{\bar m \bar n} B_{\bar m \bar n} + \bar \phi D_m
 \bar \psi^m + \phi \upeta \bar \phi}
\ee
The $\delta$-closed gauge function $\Uppsi_{\mathrm{Matter}}$ turns out to be 
$\delta$-exact: 
\be 
\Uppsi_{\mathrm{Matter}} = {\textstyle \frac{1}{(\bar \kappa \cdot
 \kappa)}}\, \delta \int_M d^4 x \sqrt{g}
\, \Scal{\bar
 \kappa_{\bar m 
 } \bar \psi_{\bar n} \bar \chi^{\bar m \bar n } + \bar \kappa^m \phi D_m \bar
 \phi }
\ee
When on computes $I_{\mathrm{Matter}}= s\Uppsi_{\mathrm{Matter}}$, one sees that $T_{ m
 n}$ and $ B_{\bar m \bar n}$ identify thereself as both scalar auxiliary fields of the Wess and Zumino multiplet.

If one combines this result with that of section \ref{yang}, one finds that
the complete supersymmetric action for a matter field coupled to the
Yang--Mills theory can be obtained by adding both gauge functions.
Let $t_\alpha$ be the generators of the Lie algebra of the
gauge group for the matter representation. The scale
factor between the Yang--Mills and the matter gauge functions can be
included in the definition of the trace. After
integration of the fields $h$, $T$ and $B$, the action is :
\begin{multline}
I_{\mathrm{YM}+\mathrm{Matter}} \approx 
\int_M d^4 x \sqrt{g} \,\Biggl( \, \trace\Scal{\frac{1}{2} F^{mn} F_{mn} +
 \frac{1}{4} \scal{ J^{m\bar n} F_{m \bar n}}^2 - \, \upchi^{mn} D_m
 \Psi_n + \upeta D^m \Psi_m } \Biggr .\\*
+ \biggl( \bar \eta D_m \bar \psi^m - \bar \chi^{\bar m \bar n} D_{\bar m}
 \bar \psi_{\bar n} - \frac{1}{2} \bar \phi \scal{ D_m D^m + D^m D_m }
 \phi \biggr .\\* - \frac{1}{2} \bar \chi_{mn} \upchi^{mn} \phi + \bar \phi \Psi_m
 \bar \psi^m - \phi \upeta \bar \eta\biggr)
 - \sum_\alpha
 \frac{1}{4 \,\trace (t^\alpha t^\alpha)} \, \Scal{\phi\, t^\alpha
 \bar \phi} \, \Scal{\phi\, t^\alpha \bar\phi} \,\, \Biggr)
\end{multline} 

The $\mathcal{N}=1$ supersymmetric action for a Yang--Mills field and a scalar complex field can therefore be directly and uniquely constructed in a twisted form, as 
an $s \delta$-exact term. 

\subsection{Embedding of the $\mathcal{N} = 1$ theory in the $\mathcal{N}=2$ theory}
\label{2to1}

Consider the complexified twisted $\mathcal{N}= 2$ theory on a
K\"{a}hler manifold. The moduli space of instantons has a
K\"{a}hler structure. The exterior differential on $M$ 
can be decomposed into Dolbeault operators and a similar property
 exists for the
BRST operator. 
 \cite{Witten} shows that the equivariant scalar BRST charge of the $\mathcal{N}=1$ theory can be identified as a component of the  scalar  BRST charge of the $\mathcal{N}=2 $ theory. Here, we  show that, starting from the $\mathcal{N}=2 $ horizontality equation, 
the projection of a general constant vector field $\kappa$ into antiholomorphic
 components  gives the $\mathcal{N}=1$ vector symmetry for the
 Yang--Mills field and the matter field. 

For a constant vector $\kappa$, with both holomorphic and antiholomorphic components, the expansion in ghost number of Eq.(\ref{nero}) determines the equivariant vector symmetry operator $\deltac$ for the $\mathcal{N}=2$ supersymmetry, that is \cite{BBT}: 
 \bea\label{neroh}
\deltac A_\mu &=& - \kappa_\mu \eta+ \kappa^\nu \chi_{\nu\mu} \CR
\deltac \Psi_\mu &=& \kappa^\nu \scal{F_{\nu\mu} - T_{\nu\mu} } +
\kappa_\mu [\Phi, \bar\Phi] \CR
\deltac \Phi &=& - \kappa^\mu \Psi_\mu \hspace{50mm} \deltac \bar\Phi = 0\CR
\deltac \eta&=& \kappa^\mu D_\mu \bar\Phi \CR
\deltac \chi_{\mu\nu} &=& - 4 \kappa_{[\mu} D_{\nu]^-} \bar\Phi \CR
\deltac T_{\mu\nu} &=& 4 \kappa_{[\mu} \scal{ D_{\nu]^-} \eta+
 \frac{1}{2} D^\sigma \chi_{\sigma|\nu]^-} - [\bar\Phi,
 \Psi_{\nu]^-}]} + 2 \kappa^\sigma D_{[\mu} \chi_{\sigma|\nu]^- } 
\eea 
Notice that, in holomorphic coordinates, the antiselfduality condition is 
$
\chi_{m\bar n} = \frac{1}{2} J_{m \bar n} J^{ n \bar m} \chi_{n \bar
 m}
$. 
Therefore, the K\"{a}hler metric allows one to define scalar fields $\chi$ and
$t$, with: 
\be 
\chi_{m\bar n} = g_{m \bar n} \chi \hspace{7mm} \chi_{\bar m n} = -
g_{\bar m n} \chi \hspace{10mm} T_{m\bar n} = g_{m \bar n} t
\hspace{7mm} T_{\bar m n} = -g_{\bar m n} t
\ee
If one chooses a constant antiholomorphic vector $\kappa^{\bar m}$, Eq.(\ref{neroh}) is projected into: \bea
\deltac A_m &=& - \kappa_m \scal{ \eta+ \chi} \hspace{40mm} \deltac
A_{\bar m} = \kappa^{\bar n} \chi_{\bar m \bar n} \CR
\deltac \Psi_m &=& \kappa^{\bar n} F_{\bar n m} + \kappa_m \scal{ t +
 [\Phi, \bar\Phi] } \quad \ \ \ \ \ \ \ \ \ \ \ 
 \deltac \Psi_{\bar m } = \kappa^{\bar n} \scal{ F_{\bar n \bar m } -
 T_{\bar n \bar m }} \CR
\deltac \Phi &=& - \kappa^{\bar m } \Psi_{\bar m} \ \ \ \ \ \ \ \ \ \ \ \ \ \ \ \ \ \ \ \ \ \ \
\ \ \ \ \ \ \ \ \ \ \ \
\deltac \bar \Phi = 0 \CR
\deltac \eta &=& \kappa^{\bar m } D_{\bar m} \bar \phi \CR
\deltac \chi_{mn} &=& - 4 \kappa_{[m} D_{n]} \bar \Phi \hspace{13mm}
\deltac \chi= - \kappa^{\bar m} D_{\bar m} \bar \Phi \hspace{13mm}
\deltac \chi_{\bar m\bar n} = 0 \CR
\deltac T_{mn} &=& \kappa^{\bar m} D_{\bar m} \chi_{mn} + 4
\kappa_{[m} \scal{ D_{n]} \eta - [\bar\Phi, \Psi_{n]}]} \CR
\deltac t &=& \kappa^{\bar m} D_{\bar m} \scal{ \eta + \chi} -
[\bar\Phi, \kappa^{\bar m } \Psi_{\bar m} ] \hspace{13mm} \deltac
T_{\bar m \bar n} = 2 \kappa^{\bar p} D_{[\bar m} \chi_{\bar p|\bar
 n]}
\eea 
By comparison with Eq.~(\ref{final}), one sees that, up to field redefinitions, the antiholomorphic component of the vector BRST symmetry of the twisted $\mathcal{N}= 2$
theory is nothing but the 
 vector 
symmetry of the $\mathcal{N}= 1$ twisted theory, as directly constructed in the last section.
 
 As for the holomorphic component of the scalar BRST operator $s$ of
 $\mathcal{N}=2$, it can be obtained by looking for an operator that
 act in a nilpotent way on the multiplet and verifies
$ \{ s_c , \deltac \} = \L_\kappa + \delta_{\mathrm{gauge}} (i_\kappa
A)$. This defines the same operator $s$ as in Eq.~(\ref{final}).

Looking for a Lagrangian $I=s_c \Uppsi$, which is invariant under antiholomorphic $s_c$ and $\deltac$ transformations, one finds both  independent 
 $\deltac$-exact topological gauge functions $\Uppsi_{\rm YM}$ and
 $\Uppsi_{\rm Matter}$ of the previous section.  Relaxing the
 antiholomorphicity condition of $\kappa$,
only a special combination of both gauge functions is $\delta_{\bar c}$ invariant, which turns out to be that of the $\mathcal{N}=2$ theory.

\subsection{Matching with the untwisted theory}
\label{untwist}
 To twist the anticommuting fields ($\Psi_m, \upeta, \upchi_{\bar m \bar n}$) of a 
topological multiplet 
into a Dirac spinor, 
a pair of covariantly constant antichiral spinors
$\zeta_{\pm}$, with $ i J^{m\bar n} \sigma_{m \bar n} \zeta_{\pm}
=\pm \zeta_{\pm}$ is needed. This implies that the manifold must be
hyperK\"{a}hler. 
We can normalize $\zeta_{\pm}$ with 
$\scal{\zeta_{-\,\aalpha}\zeta_+^{\aalpha}}=1$ . 
Then, 
 the 
 Euclidean twist formula are \cite{jaco,Witten,BS}:
 \be
\lambda_\alpha = \Psi_m {\sigma^m}_{\alpha \aalpha}
\zeta_-^{\aalpha}
\hspace{10mm}
\lambda^{\aalpha} = \upeta \zeta^{\aalpha}_+ + \upchi_{\bar
 m \bar n} {\sigma^{\bar m \bar n \aalpha}}_{\bbeta}
\zeta^{\bbeta}_+ 
\ee

Performing these changes of variables in the topological actions
found in the previous section, one finds the Euclidean ``Majorana
action'' for Dirac spinors $(\lambda_\alpha,\lambda^{\aalpha})$
described by Nicolai \cite{nico}, that is, the analytic continuation of
the Minkowski $\mathcal{N}=1$ superYang--Mills theory in Euclidean space.
(The untwisted action is independent on $\zeta_{\pm}$, as a result 
of the change of variables). Both twisted and untwisted actions do not depend on
the complex conjugates of the complex fields, i.e, $\lambda$ in the untwisted theory and 
$\upeta, \upchi$, $\Psi$ in the twisted theory. In both cases, the path integral is formally
understood as counting four real degrees of freedom. 
The Euclidean prescription must be considered as justified by 
the analytic continuation of the Minkowski case, where real
representations do exist (Majorana condition) \cite{nico}. 
 
It follows that both supersymmetric $\mathcal{N}=1$ Yang--Mills and Wess and Zumino actions are truly determined by a subsector of supersymmetry algebra with three independent generators, corresponding to both scalar and antiholomorphic vector symmetries, $s_c$ and $\delta_{\bar c}$. The closure of $s_c=
Q$ and $\delta_{\bar c}=\kappa^{\bar m} Q_{\bar m} $, and thus of the 3 generators $Q$ and $Q_{\bar m}$ under the form $Q^2=0, 
 Q_{\bar m}Q_{\bar n}+Q_{\bar n}Q_{\bar m}=0$ and 
 $ Q_{\bar m} Q+ Q Q_{\bar m}=D_{\bar m} $, stems from the property $(s + d +
\delta -i_\kappa)^2 = 0$. The above change of variables actually maps 
the four supersymmetry generators $Q^\alpha$ and $Q_{\dot \alpha}$ on the four twisted generators
 $(Q,Q_{\bar m},Q_ {mn})$. 
 The symmetry of the action under the action of the fourth twisted generator $Q_{mn}$ appears as an
additional symmetry, which is not needed, geometrically. Moreover, it is not needed for enforcing supersymmetry, for instance in technical proofs, such as those concerning renormalization.

\subsection { Formal reality condition for the action}

The Euclidean action that is the analytic continuation of the Minkowski 
 $\mathcal{N}=1$ superYang--Mills  action is not
Hermitian in the usual sense. However, as 
suggested in \cite{nico}, a ``formal complex conjugation'' can be defined, 
 for which the action is Hermitian.



Such a ``formal complex conjugation'' can be extended to the twisted case.
It is the composition of
a Wick rotation, an
ordinary complex conjugation, and an inverse Wick
rotation. The complex conjugation takes into account the fact that, in Minkowski space one has the Majorana condition for the spinors.
 This ``formal complex conjugation'', which we define as $*$, breaks the
Lorentz invariance, since we must define which direction defines the
imaginary time one. However, the operation $*$ can be covariantly
defined, by defining the temporal direction of the Minkowski space to be the covariantly
constant vector field $\kappa$ of the Euclidean manifold.
 The action of $*$ is the ordinary complex conjugation on
c-numbers and the following transformations on the fields of the theory\footnote {To
 simplify the notations, we normalize $|\kappa|$ to $1$, and define $\Lc_\xi\equiv \{i_\xi,d_A\}$.}: 
\begin{align}\label{sym}
* \partial_A *&= \bar \partial_A - g(\bar \kappa) \Lc_{\kappa - \bar
 \kappa} & * \bar \partial_A * = \partial_A +
g(\kappa) \Lc_{\kappa - \bar \kappa} \CR 
 \scal{* \Psi}_{\bar m} &= \bar \kappa_{\bar m} \upeta + \kappa^{\bar
 n} \upchi_{\bar n \bar m} \hspace{10mm} & * \upeta = \bar \kappa^m
\Psi_m \hspace{10mm} & \scal{ * \upchi}_{mn} = 2 \kappa_{[m} \Psi_{n]}
\CR
* h &= - h - \bar \kappa^m \kappa^{\bar n} F_{m \bar n} & & \CR
\scal{* \bar \psi}_m &= \kappa_m \bar \eta + \bar \kappa^n \bar \chi_{nm}
&* \bar \eta = \kappa^{\bar m} \bar \psi_{\bar m} \hspace{10mm} &
\scal{ * \bar \chi}_{\bar m 
 \bar n} = 2 \bar \kappa_{[\bar m} \bar \psi_{\bar n]} \CR 
* \phi &= - \bar \phi & * \bar \phi = - \phi \hspace{10mm} & \CR
\scal{ * T}_{\bar m \bar n} &= B_{\bar m \bar n } & \scal{*B}_{mn} =
T_{mn} \hspace{10mm}&
\end{align} 
This $*$ operation interchanges $s_c$ and
$\deltac$:
\be * s_c* = \deltac \hspace{10mm} * \deltac * = s_c \ee
The ``reality'' condition of the action means that, after integration
of auxiliary fields, one has: 
\be
* I_{\mathrm{YM} } = I_{\mathrm{YM} }
\quad
 * I_{ \mathrm{Matter}} = I_{ 
 \mathrm{Matter}}\ee 
 modulo the addition of the topological term 
 $\int_M \trace F_{\,\, \wedge} F$. In fact, this topological term is
 not invariant under the $*$ operation, only the Yang--Mills action
 $\int_M \trace F \star F$ is.

\subsection{Introduction of the WZ superpotential and Fayet--Iliopoulos term in the twisted formalism }

The introduction of a $s$ and $\delta$ invariant superpotential for 
a Calabi--Yau manifold involves terms that have zero 
ghost number only modulo 2. This parallels the breaking of
chirality induced by a superpotential in the untwisted theory. 

Consider a scalar field $\varphi $ valued in a certain representation of the gauge group. Let $f(\varphi)$ be a
superpotential, which is an analytical function of $\varphi $. Looking for
 an action $I_{\mathrm{SP}}$, which is $s$, 
 $\delta$ and $*$ invariant and has ghost number
zero modulo 2, gives: 
\begin{multline}
 \hspace{5mm} I_{\mathrm{SP}} =
 \int_M d^4 x \sqrt{g} \Biggl(
 \overline{\Omega}^{mn} \Scal{ T^A_{mn} f_A (\bar \phi) - \bar \chi^A_{mn}
 \bar \eta^B f_{AB} (\bar \phi)} \Biggr .\\* + \Omega^{\bar m \bar n}
\Scal{ B^A_{\bar m \bar n} f_A(-\phi) + \bar \psi^A_{\bar m} \bar \psi^B_{\bar
 n} f_{AB}(-\phi)} \Biggr) \hspace{10mm}
\end{multline}
 $\Omega$ and $\overline{\Omega}$ are respectively the
holomorphic and the antiholomorphic $2$-form in the Calabi--Yau
$2$-fold, and $f_A$ and $f_{AB}$ stand for the first and second derivatives of
the superpotential\footnote{The index $A$ denotes the group
 representation of the matter.}.

This term is neither $s$- nor  $\delta$-exact. However, it can be written as
follows: \be 
I_{\mathrm{SP}} = ( s + \delta ) \int_M d^4 x \sqrt{g} \Scal{
 \overline{\Omega}^{mn} 
 \bar \chi^A_{mn} f_A (\bar \phi) +2 \Omega^{\bar m \bar n} \bar
 \kappa_{\bar m} 
 \bar \psi^A_{\bar n} f_A (-\phi)}
\ee 

After expansion, one recovers, up to twist, the Wess and Zumino formula for a superpotential. The superpotential is at most a cubic
function, to ensure renormalizability.

In order to make the distinction between the topological action, which has ghost number zero, and
the superpotential, which has non vanishing even ghost number, we may
interpret the later as the insertion of an operator in the path
integral.

The $s$ and $\delta$
symmetry actually does not constrain the potential of $\phi$ (which we will name
holomorphic) to be equal to that of $\bar \phi$
(antiholomorphic): we could have chosen independent functions $f$ and
$\bar f$. However, the ``formal reality condition'' $* I_{\mathrm{SP}}
 = I_{\mathrm{SP}}$, constrains the holomorphic and the antiholomorphic 
superpotential to be related. This condition determines a real action when one goes to Minkowski space.
%

Finally, if the group has a $U(1)$ sector, one can add the Fayet--Iliopoulos term, under the
simplest invariant form: 
\be
I_{FI}=\int_M d^4 x \sqrt{g} s\, \delta\ \scal{\bar \kappa^{ m}
 A^{\scriptscriptstyle U(1) }_m }
 = 
\int_M d^4 x \sqrt{g} h^{\scriptscriptstyle U(1) }
\ee 
The integration of the auxiliary field $h$ 
gives then a mass term for the Abelian component of
the scalar fields, plus a topological term
\be
\int_M J_{\, \wedge} \, F^{\scriptscriptstyle U(1)}
\ee
 This parallels 
the property that the full superYang--Mills 
action is BRST exact, modulo a topological term. $J_{\, \wedge} \, F^{\scriptscriptstyle U(1)}$ is not invariant under the formal complex conjugation $*$. Thus,
 the formal reality condition of the twisted action implies that this topological term must be subtracted from the action. 

So, we conclude that most of the features of the supersymmetric theory in four dimensions are captured in the
TQFT formalism, from the principle of $s$ and $\delta$ invariance.

\section{Yang--Mills $\mathcal{N} = 1$ on Calabi--Yau $3$-fold}
By analogy with the four dimensional case, we might tentatively define the
horizontality condition in six dimensions, (with 8 spinorial generators in the untwisted formalism), as:
\be
(d + s + \delta - i_\kappa) \scal{ A+ c + |\kappa| \bar c } + \scal{
 A+ c+ |\kappa| \bar c}^2 = F + \Psi_{\gra{1}{0}} +
g(\kappa) \eta_{\gra{0}{0}} + i_\kappa \chi_{\gra{0}{2}} \label{compH}
\ee
However, something more elaborated
must be done, because the counting of degrees of freedom would not be quite right.
In four dimensions there is as much degrees of freedom in the
connexion, modulo gauge transformations, as in a  selfdual curvature, whereas in the six
dimensional case the counting is more subtle and involves  a scalar field \cite{BS} .

 In order to solve the question, we must refine the
K\"{a}hler decomposition of extended differentials, which include
the BRST scalar and vector operators $s$ and $\delta$. We define:
\be \t{d} = \t{\partial} + \t{\bar \partial} \ee
These extended ``heterotic'' Dolbeault operators are defined as:
\be \t{\partial} \equiv \partial
\hspace{10mm} \t{\bar \partial} 
\equiv \bar \partial + s + \delta - i_\kappa \ee
The twisted $\mathcal{N} = 1$ algebra takes into account this
asymmetry between holomorphic and antiholomorphic sectors, in a way
that generalizes the four dimensional case. In fact, for a K\"{a}hler
manifold, the Bianchi identity of the curvature $F$ can be decomposed
as follows: 
\be 
\partial_A F_{\gra{2}{0}} = 0 \hspace{7mm} \partial_A F_{\gra{1}{1}} +
\bar \partial_A F_{\gra{2}{0}} = 0 \hspace{7mm} \bar \partial_A
F_{\gra{1}{1}} + \partial_A F_{\gra{0}{2}} = 0 \hspace{7mm} \bar
\partial_A F_{\gra{0}{2}} = 0
\ee
 Eq.~(\ref{compH}) decomposes into two
horizontality conditions:
\bea
(\bar \partial + s + \delta - i_\kappa ) \scal{ A_{\gra{0}{1}} + c +
 |\kappa| \bar c} +
\scal{ A_{\gra{0}{1}} + c + |\kappa| \bar c}^2 &=& 
F_{\gra{0}{2}} + i_\kappa \chi\CR
\bigl\{\bar \partial_A + s_c + \deltac - i_\kappa , \partial_A \bigr\} &=&
F_{\gra{1}{1}} + \Psi + g(\kappa) \eta
\eea
with their Bianchi identities 
\bea
(\bar \partial_A + s_c + \deltac - i_\kappa) \scal{ F_{\gra{0}{2}} +
 i_\kappa \chi} &=& 0 \CR
(\bar \partial_A + s_c + \deltac - i_\kappa) \scal{ F_{\gra{1}{1}} +
 \Psi + g(\kappa) \eta } + \partial_A \scal{ F_{\gra{0}{2}} +
 i_\kappa \chi} &=& 0
\eea
These equations only determine part of the BRST algebra. Indeed, they miss a dependence on 
 the holomorphic curvature
$F_{\gra{2}{0}}$. One must therefore introduce a third horizontality
condition, which completes Eq.~(\ref{compH}), and enforces the definition of $F_{\gra{2}{0}}$ as a curvature. In order to
have the necessary balance between the ghosts and antighosts degrees of
freedom, we understand that one further degree of freedom with ghost
number one must be introduced. It can be represented as a
holomorphic $3$-form $\varsigma$. The third horizontality condition and Bianchi identity are: 
\bea
(\bar \partial_A + s_c + \deltac - i_\kappa) \, \varsigma_{\gra{3}{0}} = \bar
\partial_A \varsigma_{\gra{3}{0}} + \Phi_{\gra{3}{0}} + g(\kappa) F_{\gra{2}{0}} 
\\
(\bar \partial_A + s_c + \deltac - i_\kappa) \scal{ \bar
\partial_A \varsigma + \Phi + g(\kappa) F_{\gra{2}{0}}} = [
F_{\gra{0}{2}} + i_\kappa \chi,\, \varsigma \,]
\eea
Expending the equations in form degree and ghost number determines the action of $s_c $ and $\deltac$:
\be \begin{array}{rclcrcl}
s_c A_m &=& \Psi_m &\hspace{10mm}& \deltac A_m &=& \kappa_m \eta \CR
s_c A_{\bar m} &=& 0 & & \deltac A_{\bar m} &=& \kappa^{\bar n}
\chi_{\bar n \bar m} \CR
s_c \Psi_m &=& 0 & & \deltac \Psi_m &=& \kappa^{\bar n} F_{\bar n m} -
\kappa_m h \CR
s_c \eta &=& h & & \deltac \eta &=& 0 \CR
s_c \chi_{\bar m \bar n} &=& F_{\bar m \bar n} & & \deltac \chi_{\bar
 m\bar n} &=& \kappa^{\bar p} \bar \Phi_{\bar p \bar m \bar n} \CR
s_c h &=& 0 & & \deltac h &=& \kappa^{\bar m} D_{\bar m} \eta \CR
s_c \bar \Phi_{\bar p\bar m \bar n} &=& 3 D_{[\bar p} \chi_{\bar m
 \bar n]} & & \deltac \bar \Phi_{\bar p \bar m \bar n} &=& 0 \CR
s_c \varsigma_{pmn} &=& \Phi_{pmn} & & \deltac \varsigma_{pmn} &=& 3
\kappa_{[p} F_{mn]} \CR
s_c \Phi_{pmn} &=& 0 & & \deltac \Phi_{pmn} &=& \kappa^{\bar q} D_{\bar
 q} \varsigma_{pmn} - 6 \kappa_{[p} D_m \Psi_{n]}
\end{array} \ee
The $\delta$ invariance determines the following topological gauge function:
\be
\Uppsi = \int_M d^6 x \sqrt{g}\, \trace \Scal{ \frac{1}{2} \chi^{mn}
 F_{mn} + \eta \scal{ h + i J^{m\bar n} F_{m \bar n}} - \frac{1}{6}
 \bar \Phi^{mnp} \varsigma_{mnp}} 
\ee
 Integrating out $h$, 
$\Phi$, $\bar\Phi$ gives the following $s$ and $\delta$ invariant action:
\be
I= s \Uppsi =\int_M d^6 x \sqrt{g} \,\trace \Biggl( \, \frac{1}{2} F^{mn}F_{mn} +
\frac{1}{4} \scal{ J^{m\bar n} F_{m\bar n}}^2 \Biggr.
- \chi^{mn} D_m \Psi_n + \eta D_{\bar m } \Psi^{\bar m} + \frac{1}{2}
\varsigma^{\bar p \bar m \bar n} D_{\bar p} \chi_{\bar m \bar n}
\Biggr)
\ee
This action is the twisted, $\mathcal{N}=1,D=6$
supersymmetric action, up to a topological term
$
\int_M\, i J_{\, \wedge} \trace F_{\, \wedge} F
$, as in \cite{BS}.
As in the four dimensional case, this action is $s \delta$-exact, 
\be
I=s\, \delta { \textstyle \frac{1}{(\kappa \cdot \bar \kappa)}} \int_M
d^4 x \sqrt{g} \, 
\trace \biggl( \eta \bar \kappa^m \Psi_m + \frac{1}{2}
\varsigma_{pmn} \bar\kappa^p \chi^{mn} 
+ \bar \kappa^{[m} g^{n]\bar m} \scal{ 3 A_{[\bar m} \partial_m
 A_{n]}
+ 2 A_{[\bar m} A_m A_{n]}}
\biggr) 
\ee
The last term is just 
$i g(\bar \kappa)_{\, \wedge} J_{\, \wedge} \trace \scal{ A d A +
 \frac{2}{3} A ^3} $.
 

\section{Conclusion}
Putting together the results of this paper and of \cite{BBT}, we
reach an interesting  conclusion for the super Yang--Mills
symmetries with 4, 8 and 16 generators, which can be
represented as $\mathcal{N}=1$ theories in 4 and 6 dimensions and
$\mathcal{N}=2$ theories in 4 and 8 dimensions.

On the one hand, one can directly see that the known spinorial
generators of the superPoincar\'e ``on-shell algebra'' can be mapped
on tensor operators with either Lorentz indices or holomorphic
indices,   as follows: \be \begin{array}{rclcrclrclcrcl}
\mathcal{N} =2 & D=4,8 & &(Q_{ },&Q_{\mu},&Q_{\mu\nu^-})& & & &
& & & \CR 

\mathcal{N}=1& D=4 & &(Q_{ },&Q_{\bar m,}&Q_{mn})& & &
& & & & \CR 
\mathcal{N}=1 & D=6 & &(Q_{ },&Q_{\bar m},&Q_{mn},& Q_{\bar m\bar
 n\bar p})& & & & & & \CR
 \end{array} \ee
In some cases,  auxiliary fields exist, giving ``off-shell'' closed transformations.

 On the other hand, we found a reverse construction, that  clarifies
the structure of supersymmetry. 
 In all cases, the set of both scalar and
 vector generators $(Q,Q_{\mu})$ or $(Q,Q_{\bar m})$ are 
 determined by  horizontality conditions, which  only involve fields
 related to the geometry of the Yang--Mills fields. The theory is in
 fact defined by these equivariant scalar and vector BRST differential
 operators. This reduced set 
 of generators builds a closed ``off-shell'' algebra, which is large
 enough to determine the supersymmetric actions. It allows one to
 reconstruct by an untwisting procedure the complete structure of 
 Poincar\'e supersymmetry.
 Moreover, the dependence on Faddeev--Popov ghosts can be computed, in a way that clearly allows a consistent and convenient covariant Yang--Mills gauge-fixing of the supersymmetric actions.
 Tensor generators, such as $Q_{\mu\nu^-}$, $Q_{ m n}$,
 $Q_{\bar m\bar
 n\bar p}$, are decoupled sets of generators. They appear as
additional symmetries of the actions that are defined by invariance
under  scalar and vector topological symmetries, but are not needed,
neither for geometrical reasons, nor for the sake of defining the
theories. However, by completion, they allow  the construction of the
irreducible set of on-shell supersymmetry spinorial generators in flat
space. 
 

\subsection*{Acknowledgments}

This work was partially supported under the contract ANR(CNRS-USAR) \\ \texttt{no.05-BLAN-0079-01}.


\end{document}